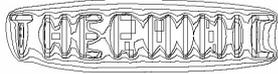

# A MORE FLEXIBLE REALIZATION OF THE SUNRED ALGORITHM

*László Pohl, Vladimír Székely*

pohl@eet.bme.hu, szekely@eet.bme.hu
Budapest University of Technology and Economics
Department of Electron Devices
1521 Budapest, Hungary

**ABSTRACT**

The high dissipation of integrated circuits means serious problems for packaging and for the design of complex electronic systems. Another important area of research and development nowadays is the integration of sensors and micromechanical systems (MEMS) with electronic circuits. The original Successive Node Reduction (SUNRED) algorithm handles well the first area but require revision for electro-thermal or mechanical fields. As a first stage the updated algorithm is able to solve thermal fields as the original, but with the application of flexible boundary connection handling, it can be much faster than the original. By using object-oriented program model the algorithm can handle non-rectangular 3D fields, and SUNRED mesh resolution is arbitrary, not have to be the power of two anymore.

## 1. INTRODUCTION

The Successive Node Reduction (SUNRED) algorithm is a solution method for Finite Differences Method [1] models. The algorithm works in 2D [2] and in 3D [4]. The original algorithm has been developed to solve thermal problems but it has been extended for electro-static fields [3,5]. Section 2 and 3 gives a brief overview of the SUNRED model and algorithm.

This algorithm is faster than FEM programs with the same resolution [4], but it is still a time-consuming process. There are two ways for further reduction of computation time without losing precision: optimizing the code and reducing the number of nodes of the model electrical network.

With code optimizations the solver program can be 1.5-2 times faster than the original. These techniques are commonly known (e.g. [6]), they are not subjects of the article.

Reducing the node number can provide much higher solution speed. Number of floating-point operations in SUNRED 3D is proportional to $P^2$ where P is the number of nodes [4]. The original algorithm is able to use non-equidistant grid which means fine grid where it is crucial and raw grid where high accuracy is not needed.

In this paper we step further. Section 4 presents flexible boundary node handling. When the boundary conditions are fixed during a simulation or simulation sequence, the boundary conditions can be integrated into the structure. In this case the boundary nodes are hidden during the node reduction process so the computation time decreases. If the boundary nodes are non-integrated, non-hidden, the change of boundary conditions not require new node reduction, only a much faster process (~P). In the new algorithm the user can decide which boundary nodes are integrated or non-integrated.

Another possibility for the node number reduction is the using of flexible cell structure, this technique is discussed in section 5. The original algorithm is restricted to a $2^n \times 2^n \times 2^m$ resolution rectangular prism (similar to Fig.1.a), in the new solver there are no such restrictions.

Section 6 gives an example which compares the solution times when a complex structure is analyzed and the investigated field is badly fits into a rectangular prism.

## 2. SUNRED MODEL

The model of thermal or electro-static field (Fig.1.a) is an electrical network (Fig.1.b). The purpose of the simulation is to determine the temperature/voltage of the nodes in the center of the cells. The boundary conditions are known.

In the first step the following equation is determined for each cell:

$$\underline{I} = \underline{\underline{Y}}\,\underline{U} + \underline{J} \qquad (1)$$

I and U are the unknown current and voltage vectors of the external nodes of the cell; Y and J are the admittance matrix and inhomogeneous current vector, these are determined from the field parameters. I or U

*L. Pohl, V. Székely*
*A more flexible realization of the SUNRED algorithm*

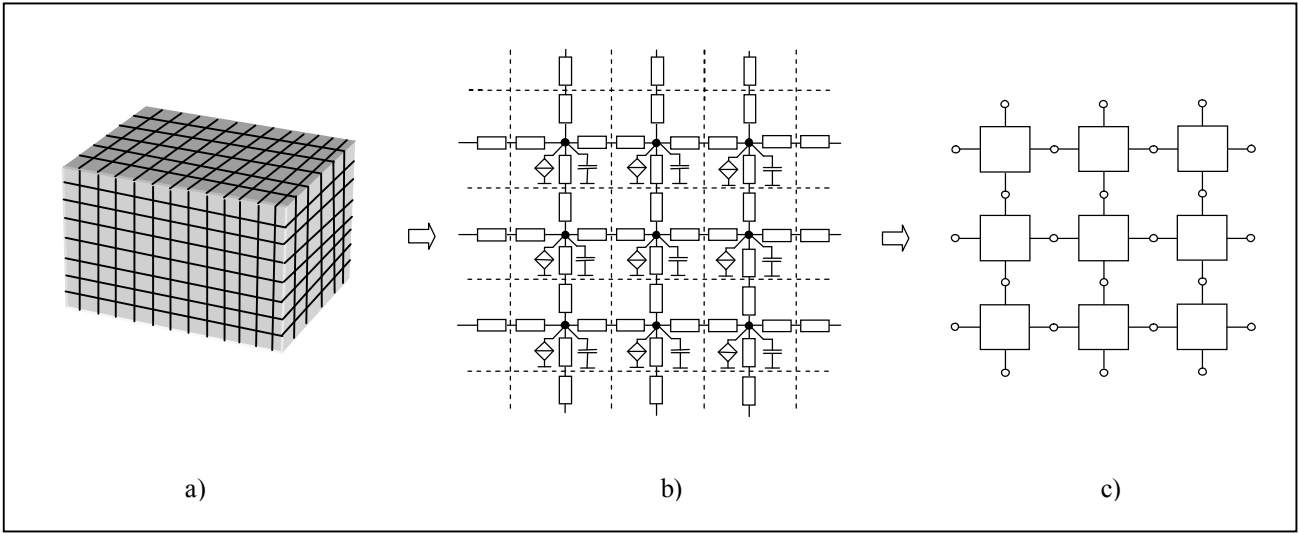

Fig.1 a) 3D rectangular field   b) SUNRED version of Finite Differences model in 2D   b) Reduced first level cells values are known only on the boundaries of the simulated field, given by the boundary conditions. The Successive Node Reduction algorithm can calculate U and I values from Y, J and boundary conditions.

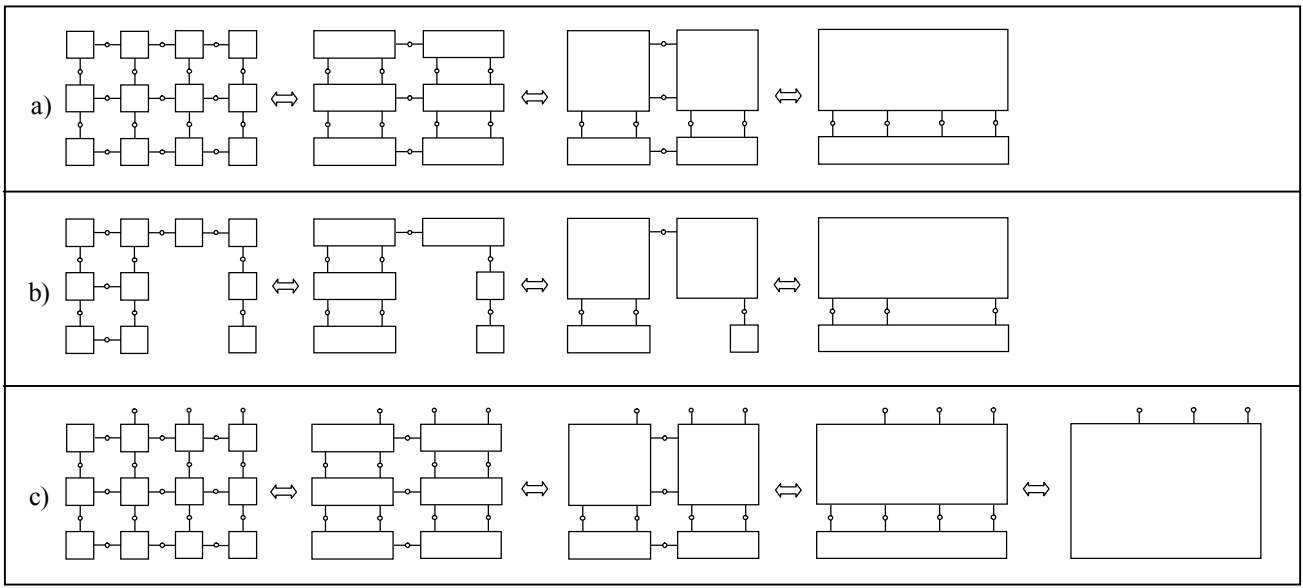

Fig.2 Successive node reduction with a) rectangular field b) non-rectangular field c) non-integrated external nodes

Equation (1) describes each cell properly, so the internal structure of the cell is not important during the calculations which means the internal node of the cell can be "reduced" (Fig.1.c), all cells will be described by (1).

### 3. NODE REDUCTION

Fig 2 a) and b) shows the steps of successive node reduction when boundary nodes are integrated, Fig.2.c) is the general case. From left to right in each step the cells are merged which means the reduction of the nodes between the merged cells i.e. the calculation of Y matrices and J vectors for the new cells. The equations of the reduction are published e.g. in [3] or [5].

Finally we gain one cell this cell have only boundary nodes, which means equation (1) can be calculated for this cell, because the boundary condition determines the U or the I value for each node. The solution of a linear equation system provides all U and I values for the boundary nodes. In a successive backward substitution process (Fig.2. right to left) the U and I values of the internal nodes are calculated.

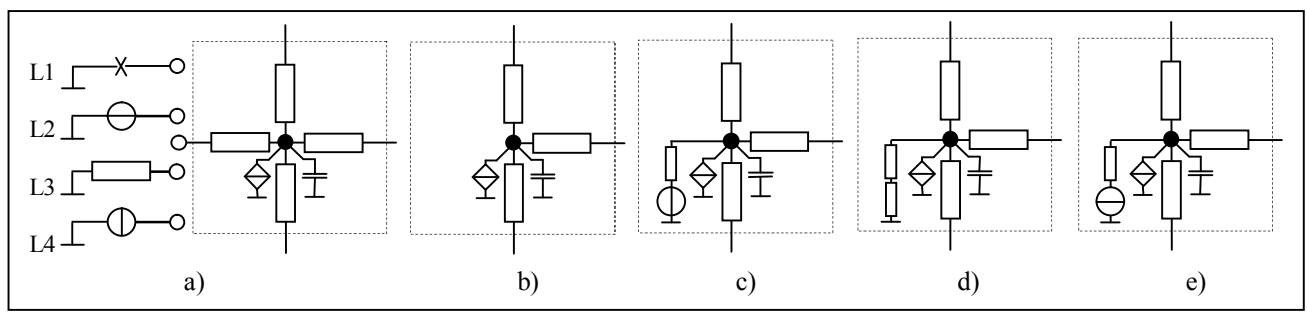

Fig.3. Boundary conditions (BC). a) Boundary cell with the four types of BCs. b)-e) Integrated BCs

| | Grid resolution | Original solver | New solver |
|---|---|---|---|
| Thermal DC analysis running time (2.8 GHz Intel Pentium 4C processor and 1 GB's of memory) | 512×512×1 | 21.2 s | 19.3 s |
| | 256×256×2 | 19.1 s | 9.1 s |
| | 128×128×4 | 18.4 s | 6.7 s |
| | 64×64×8 | 18.0 s | 5.1 s |
| | 32×32×16 | 18.0 s | 3.5 s |
| | 183×67×6 | N/A | 16.1 s |

Table.1 Computation time for the same structure with partially integrated boundary conditions (original solver) and fully integrated boundary conditions (new solver)

| Grid resolution | x×y | 32×32, 64×64, …, 1024×1024 | m×n×o  m, n, o ∈ positive integer |
|---|---|---|---|
| | z | 1, 2, 4, 8, 16 | |

Table.2. Supported resolutions (new solver: rectangular mode)

## 4. INTEGRATION OF BOUNDARY NODES, EXTERNAL NODES

The computation time highly depends on the number of nodes of the simulated field: number of necessary floating point operations in 3D is Ordo($P^2$), P is the full number of the nodes of the structure [4]. The computation time can be reduced by the reduction of the number of nodes.

One possibility to reduce the number of nodes is the integration of boundary conditions. Fig.3 a) shows the four supported types of boundary conditions of SUNRED, and a boundary cell. The integration of the boundary condition means that the terminating two-pole is handled as part of the cell, so the cell will have fewer external nodes (Fig.3 b-e).

Advantage of integration:
- faster solution in most cases

Advantages of non-integration:
- When the simulated field does not change between simulations, but the boundary conditions do and the boundary nodes stay external, there is no need to repeat successive node reduction. Applying new boundary conditions on external nodes and recalculate the voltages of internal nodes is extremely fast compared to a full reduction process (Ordo(P)).
- In some cases not typical boundary conditions are used. E.g. two, separately solved fields can be connected through boundary nodes [5], or compact models can be join to the field. These applications need external nodes.

In the original SUNRED solver it is fixed that which boundary nodes are integrated or non-integrated: top and bottom side nodes are integrated, east, west, north and south nodes are non-integrated. In this case the computation time is acceptable and the program structure is simple and controllable.

In the new SUNRED solver the distribution of integrated and non-integrated boundary nodes is totally flexible. In Fig 2.c) there are three non-integrated nodes, Fig.2.a) shows a fully integrated structure. Moreover in the new solver all internal (non-boundary) nodes of the structure can be handled as "external", not only the boundary nodes.

When fully integrated structure is used, the last reduction step is missing; in this case the full current of

*L. Pohl, V. Székely*
*A more flexible realization of the SUNRED algorithm*

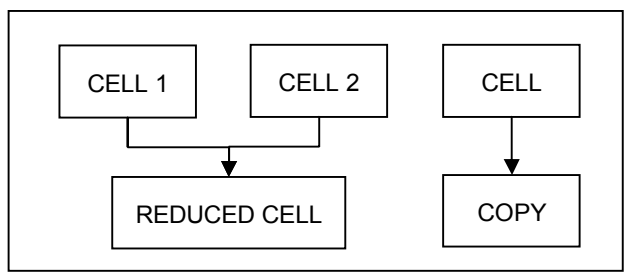

Fig.4. The two way of creation of the next level cell

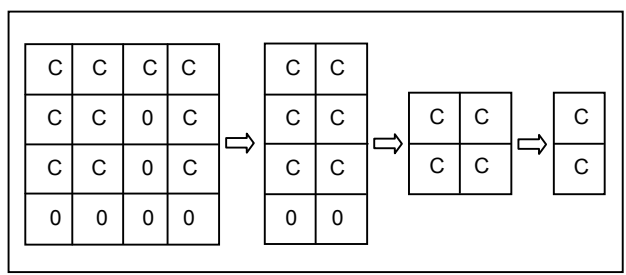

Fig.5 Cell pointer array during the reduction

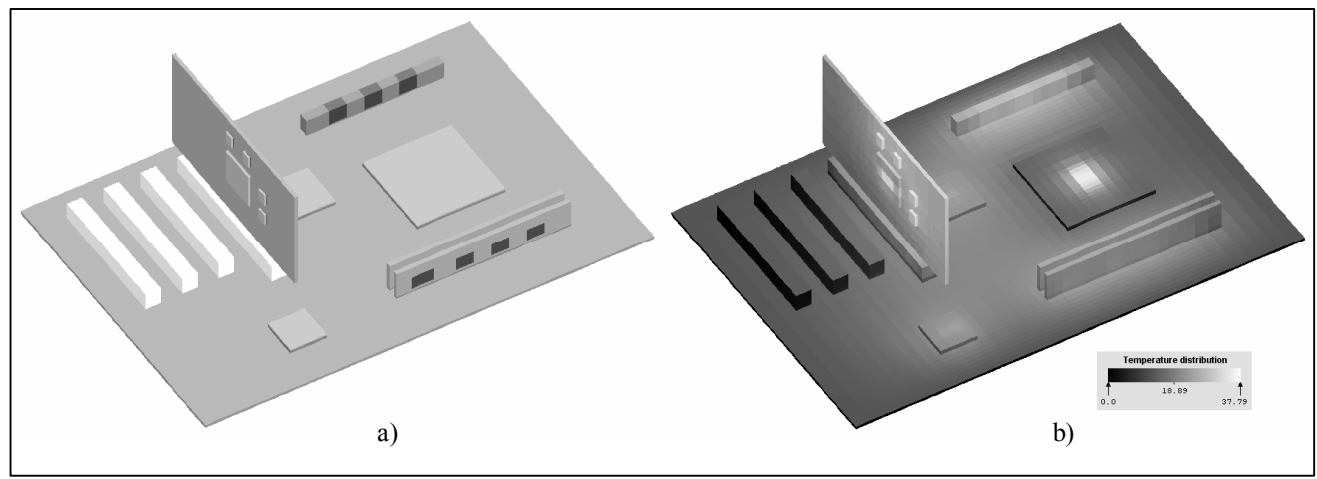

a)  b)

Fig.6. a) SUNRED model of a motherboard  b) Simulation result: temperature distribution

the internal nodes is the inhomogeneous current (J), so the voltage of the remaining nodes of the last step can be calculated as follows:

$$\underline{U} = -\left(\underline{\underline{Y}}_1 + \underline{\underline{Y}}_2\right)^{-1}\left(\underline{J}_1 + \underline{J}_2\right) \qquad (2)$$

Where U is the voltage vector of the last-step internal nodes, $Y_1$ and $J_1$ is the admittance matrix and inhomogeneous current vector of the one last cell (Fig.2.a upper cell), $Y_2$ and $J_2$ is admittance matrix and inhomogeneous current vector of the other last cell (Fig.2.a lower cell).

Table.1. compares the computation time of the same structures by the original solver and by the new solver (new solver in fully integrated mode). At multiple layer structures the difference is very high: a half order of magnitude. Note that the main cause of this difference is that the external nodes are never reduced during the node reduction, and in the final steps the cells will have much more nodes than in integrated case.

**5. FLEXIBLE CELL-STRUCTURE**

The new solver has been written in object-oriented C++, where every cell is an object that can administrate itself.

The cells store their Y matrix, J, U, I vectors and other subsidiary data. In these cells the nodes of the different sides are handled separately. Thank for the C++ the code is controllable. The original solver has been written in C where this administration would have been very difficult.

The cell administration is the simplest in that case when all cells have the same structure (the same number of nodes on the same side). The original solver has followed this model. In this case the resolution has to be the power of two, in all direction the same. The original solver used a trick, so only the x-y direction has to be the same size, z can be different, see Table.2. The explanation for the power-of-two can be seen in Fig.2 a) or c): the third step has resulted two types of cells, which contradict the requirement of identical cell structures.

The new solver uses a power of two sized pointer array to store cells, but where there is no cell, a NULL pointer is stored (Fig.5). This let us not only the application of non-power-of-two sized resolution rectangular fields, but also non-rectangular fields (Fig.2.b). The reduction method can get one or two cells (Fig.4). If it obtains two cells, the new cell reduces the common nodes, and merges the cells into one in it. If it obtains one cell, the new one simply copies that into it.

Two NULL pointers result a NULL pointer in the next level pointer array.

The application of non-power-of-two resolution grid can speed up the simulation because we do not have to use higher resolution i.e. more nodes than it is needed, and fewer node number means lower computation time.

## 6. APPLICATION EXAMPLE

As an example, we have chosen a complex system: a motherboard with processor, RAM and a video card (Fig.6.a). This field fits badly into a rectangular prism so if the original solver has been used which not supports non-rectangular fields, a lot of extra time would be needed for the calculation.

The system contains the following dissipating elements: CPU (65W), Voltage Regulator Module (10W), North Bridge (5W), South Bridge (8W), two RAM modules (5W), GPU (25W), Video RAM (10W).

The heat sinks and coolers were modeled as different HTC boundary conditions: CPU (2000 W/m$^2$K), VRM and NB (300 W/m$^2$K), SB (500 W/m$^2$K), GPU and VRAM (1500 W/m$^2$K). RAM modules were not cooled.

We have accomplished the simulation in two ways by the new solver. First the model was a 64×64×16 grid resolution rectangular field (like Fig.1.a). In this case the volume between Fig.1.a) and Fig.6.a) was filled with "air". In the second case the model was identical with Fig.6.a), and everywhere where in the first model the air contacted the surface we used HTC= 2 W/m$^2$K boundary termination.

After DC simulation we have got the temperature distribution of the structure. The differences between the results of the two models were minimal (<0,5%). The hottest components were the CPU, GPU and VRAM (30.1-37.5 °C temperature rise). The other components: VRM (16.5-24.4 °C); NB, SB and RAM (12.6-18.0 °C).

The most important difference between the two simulations was the computation time. In the first case the analysis took 35 sec, in the second case it took only 0.93 sec.

The original solver finishes the analysis of a similar, 64×64×16 resolution structure in 142 sec, but it cannot solve this problem because it not supports non-rectangular structures.

## 7. CONCLUSION

The development of Successive Node Reduction algorithm has resulted a faster and more flexible solver.

First group of changes has touched the external node handling. In the new solver the boundary nodes can be external nodes, or they can be integrated, and the internal nodes can be changed to external. By using external nodes external excitations or networks can be coupled to the field, or separately solved fields can be joined. By integrating boundary conditions the solution can be much faster.

Second main direction of changes means flexible cell structure. By using self-administrating cells and cell pointer arrays the finite differences grid resolutions can be freely chosen and non-rectangular structures can be used.

The next purpose of development is electro-thermal simulation by successive node reduction.

## 8. REFERENCES

[1] G. Beer and J. O Watson: Introduction to finite and boundary element methods, *John Wiley & Sons*, Chichester, 1992

[2] V. Székely: SUNRED a new thermal simulator and typical applications, *3rd THERMINIC Workshop*, 21-23 September, Cannes, France, pp. 84-90, 1997

[3] V. Székely, M. Rencz: Fast field solvers for thermal and electrostatic analysis, *DATE'98*, 23-26 February, Paris, France, Proceedings pp. 518-523, 1998

[4] A. Páhi, V. Székely, M. Rosenthal, M. Rencz: 3D extension of the SUNRED field solver, *4th THERMINIC Workshop*, 27-29 September, Cannes, France, pp. 185-190, 1998

[5] L. Pohl, V. Székely: Developments of the SUNRED algorithm, *9th THERMINIC Workshop*, Sept. 24-26, Aix-en-Provence, France, Proceedings pp. 197-200, 2003

[6] Streaming SIMD Extensions - Matrix Multiplication, *Intel Corporation*, June 1999

[7] M. Furmañczyk, A. Napieralski, E. Yu, A. Przekwas, M. Turowski: Thermal Model Reduction for IC packages and MCM's, *4th Therminic Workshop*, Sept. 27-29, Cannes, France, pp 135-138, 1998